\documentclass[aps,pra,a4paper,showpacs,twocolumn,floatfix]{revtex4}
\usepackage[latin1]{inputenc}
\usepackage{amsfonts}
\usepackage{amssymb}
\usepackage{amsmath}
\usepackage{calc}
\usepackage{graphicx}
\usepackage{epsfig}
\usepackage{bm}
\usepackage{ulem}
\usepackage{setspace}
\usepackage{subfigure}

\newcommand{\ket}[1]{|#1\rangle}
\newcommand{\bra}[1]{\langle#1}

\begin{document}
%%%%%%%%%%%%%%%%%%%%%%%%%%%%%%%%%%%%%%%%%%%%%%%%
%%%%%%%%%%%%%%%%%%%%%%%%%%%%%%%%%%%%%%%%%%%%%%%%

\title{Analysis of a quantum memory for photons \\
based on controlled reversible inhomogeneous broadening}

\date{\today}
\pacs{03.67.Hk, 32.80.-t, 42.50.Md}
\author{Nicolas Sangouard}
\altaffiliation{Also at: Laboratoire de Spectrom\'etrie Physique, CNRS-Universit\'e de Grenoble 1, St. Martin d'{}H\`eres, France}
\affiliation{Group of applied Physics-Optics, University of Geneva, Switzerland}
\author{Christoph Simon}
\affiliation{Group of applied Physics-Optics, University of Geneva, Switzerland}
\author{Mikael Afzelius}
\affiliation{Group of applied Physics-Optics, University of Geneva, Switzerland}
\author{Nicolas Gisin}
\homepage{http://www.gap-optique.unige.ch}
\affiliation{Group of applied Physics-Optics, University of Geneva, Switzerland}

\begin{abstract}
We present a detailed analysis of a quantum memory for
photons based on controlled and reversible inhomogeneous
broadening (CRIB). The explicit solution of the equations of
motion is obtained in the weak excitation regime, making it possible to
gain insight into the dependence of the memory efficiency on the
optical depth, and on the width and shape of the atomic
spectral distributions. We also study a simplified memory
protocol which does not require any optical control fields.
\end{abstract}

\maketitle
%\tableofcontents
%\newpage

%%%%%%%%%%%%%%%%%%%%%%%%%%%%%%%%%%%%%%%%%%%%%%%%%
%%%%%%%%%%%%%%%%%%%%%%%%%%%%%%%%%%%%%%%%%%%%%%%%%
\section{Introduction}

The implementation of quantum memories for photons is an
important goal in quantum information processing. It would
allow the realization of on-demand single-photon sources
based on heralded sources \cite{Hong86, Fasel04} and provide a basic
ingredient for quantum repeaters \cite{Briegel98}.
Several different approaches to the
realization of such memories have been proposed, using both
single absorbers in high-finesse cavities \cite{Cirac97}
and dense atomic ensembles. The latter proposals include
the use of off-resonant interactions \cite{Polzik}, of
electromagnetically induced transparency (EIT)
\cite{Fleischhauer02}, and of non-standard photon echoes
\cite{Moiseev01}. On the experimental side, storage and
retrieval of classical light has been realized using EIT
\cite{EIT} and photon echoes \cite{Mossberg}. 
Coherent states of light have been stored in an atomic
ensemble with higher-than-classical fidelity using off-resonant interactions \cite{Julsgaard04}. Recently the storage and retrieval of single photons 
have been reported using EIT in atomic ensembles \cite{Chaneliere05, Eisaman05}.

In the present work we are following the approach that
originated in Ref. \cite{Moiseev01}, where it was shown
that highly efficient photon echoes \cite{Abella64} can be
generated in a gas of atoms exploiting the fact that the
Doppler shift changes sign if the propagation direction of
the light is reversed. A first adaptation of the effect to
solid state systems was suggested in \cite{Moiseev03} using
nuclear magnetic resonance. Refs. \cite{Nilsson05} and
\cite{Kraus06} proposed an attractive experimental
realization of the same principle using controlled
reversible inhomogeneous broadening (CRIB). These proposals
combine spectral hole burning techniques and the use of
controllable Stark shifts from electric field gradients.
First proof-of-principle experimental demonstrations of the
CRIB approach have recently been performed
\cite{Alexander06, Hetet06}.

Here we perform a detailed theoretical analysis of the CRIB 
quantum memory protocol for finite optical
depth and general atomic distributions. In section
\ref{principle} we give the equations of motion for the
full system atoms plus light and recall the principle of a
memory based on CRIB. In section \ref{resolution} we derive
the general solution of the equations of motion under the
condition of weak excitation. We discuss both the complete
memory protocol, where the output field is emitted in the
backward direction, and a simplified protocol, which does
not use any optical control fields, leading to forward
emission. In section \ref{finite_optical_depth} we study
the dependence of the memory efficiency on the optical
depth of the medium. We show that it approaches one for the
complete protocol for sufficiently large optical depth,
whereas it can be over 50 \% for the simplified protocol,
for which the efficiency is limited by reabsorption. In
section \ref{discussion} we show that the width of the
initial atomic distribution determines the possible storage
time. We furthermore analyze what is the optimal broadening
for a given initial distribution and pulse width. Finally
we investigate the effect of shapes of the atomic
distribution and light pulse. Section \ref{conclusion} gives our
conclusions.
%%%%%%%%%%%%%%%%%%%%%%%%%%%%%%%%%%%%%%%%%%%%%%%%%%
\section{General principle}\label{principle}
A light pulse propagates through a medium composed of two-level atoms. It resonantly couples the two atomic states, the ground state $\ket{g}$ and the excited state $\ket{e}.$ We look at the behavior of the positive frequency part of the slowly time-varying envelope $E(z,t)$ of the light field decomposed in forward and backward modes
\begin{equation}
E(z,t)= E_{\rm{f}}(z,t)e^{i \omega_0z/c}+E_{\rm{b}}(z,t)e^{-i \omega_0z/c}
\end{equation}
in a one-dimensional light propagation model. This one-dimensional model is well adapted to the propagation either in an optical wave guide or in a bulk medium under the condition that $A/\lambda  \ell \gg 1,$ $A$ being the area of the light beam, $\lambda$ the wavelength and $\ell$ the propagation length. The central frequency $\omega_0$ of the light pulse is detuned from the atomic transition frequency $\omega_{\rm eg}$ by $\Delta:=\omega_0-\omega_{\rm eg}.$ The atomic transitions undergo an inhomogeneous broadening that we consider uniform in space such that the density of atoms associated to the detuning $\Delta$ is $\rho(\Delta)$ \cite{symmetry_density}. To describe the properties of the atomic ensemble, we define a mean field per atoms, slowly varying in time
\begin{equation}
\sigma_{\rm ij}(z,t;\Delta):= \frac{1}{N(\Delta,z)} \sum_{n=1}^{N(\Delta,z)} \ket{i}_{nn} \bra{j}|,
\end{equation}
${\rm i,j}$ standing for ${\rm e}$ or $\rm{g}.$ In the above sum, the atom index $n$ runs over all atoms $N(\Delta,z):=\rho(\Delta)\delta z\delta\Delta/L$ with detuning within the interval $[\Delta-\delta\Delta/2,\Delta+\delta\Delta/2]$ and position within the interval $[z-\delta z/2,z+\delta z/2].$ $L$ denotes the length of the medium.
%%%%%%%%%%%%Atoms
The Hamiltonian describing the system atoms plus light is given in the rotating wave approximation by
\begin{eqnarray}
H=\int_{-\infty}^{+\infty} d\Delta \frac{\rho(\Delta)}{L} \int_{0}^{L} dz \big[\Delta \sigma_{\rm ee}(z,t;\Delta)&& \\
-\wp E(z,t)\sigma_{\rm{eg}} \left(z,t;\Delta\right)&+&\hbox{H.c.}\big] \nonumber
\end{eqnarray}
$\wp$ being the dipole moment of the transition $\ket{e}$-$\ket{g}.$ In analogy with the light field, the positive frequency part of the operator associated to the atomic coherence $\sigma_{\rm ge}$ is decomposed into two counter-propagating contributions
\begin{equation}
\label{sigma}
\sigma_{\rm ge}(z,t;\Delta)=\sigma_{\rm{f}}(z,t;\Delta)e^{i\omega_0z/c}+\sigma_{\rm{b}}(z,t;\Delta)e^{-i\omega_0z/c}.
\end{equation}
Under the approximation that most of the population stays in the ground state $\sigma_{\rm gg} \approx 1,$ which is well justified for single or few photons light storage, the evolution of the atomic coherence is given by the following Heisenberg-Langevin equations
\begin{subequations}
\label{atomic_evolution}
\begin{equation}
\label{atomforward_motion}
\frac{\partial}{\partial t}{\sigma}_{\rm f}(z,t;\Delta)=-i\Delta \sigma_{\rm f}(z,t;\Delta)+i\wp E_{\rm f}(z,t),
\end{equation}
\begin{equation}
\label{atombackward_motion}
\frac{\partial}{\partial t}{\sigma}_{\rm b}(z,t;\Delta)=-i\Delta \sigma_{\rm b}(z,t;\Delta)+i \wp E_{\rm b}(z,t).
\end{equation}
\end{subequations}
Here we neglect the homogeneous decoherence. We will however take into account inhomogeneous dephasing in the following, cf. section \ref{memo_eff}.
%%%%%%%%%%%%%%Light
Using the slowly varying approximation, the evolution of the forward and backward components of the light pulse is given by
\begin{subequations}
\label{light_evolution}
\begin{equation}
\label{pulseforward_motion}
\left(\frac{\partial}{\partial t} + c\frac{\partial}{\partial z}\right)E_{\rm{f}}(z,t)=i \beta \int_{-\infty}^{+\infty} d\Delta \-\ G(\Delta) \sigma_{\rm f}(z,t;\Delta),
\end{equation}
\begin{equation}
\label{pulsebackward_motion}
\left(\frac{\partial}{\partial t} - c\frac{\partial}{\partial z}\right) E_{\rm{b}}(z,t)= i \beta \int_{-\infty}^{+\infty} d\Delta \-\ G(\Delta) \sigma_{\rm b}(z,t;\Delta)
\end{equation}
\end{subequations}
where $\beta$ is defined by $\beta:=g_0^2N \wp$ with $g_0:=\sqrt{\omega_0/(2 \epsilon_0 V)}.$ $N:=\int d\Delta \rho(\Delta)$ is the number of atoms in the quantization volume $V$ such that $\rho(\Delta)=N \times G(\Delta)$ where $G(\Delta)$ is the normalized spectral atomic distribution.\\
A complete description of the system is given by the set of Eqns. (\ref{atomic_evolution})-(\ref{light_evolution}). These equations of motion describe a situation in which the presence of the laser field introduces changes in the medium by inducing a polarization of the atomic ensemble. Reciprocally, the atomic polarization can serve as a source for the optical field. Owing to the consideration of weak excitations, the coupled equations are linear and of first order offering the possibility to solve them analytically. Furthermore, the linearity of the equations of motion implies the equivalence between classical and single-photon dynamics. In the following, the quantities $E_{\rm b}$ and $E_{\rm f}$ can be interpreted either as classical fields or as single-photon wave-functions. Analogously, $\sigma_{\rm b}$ and $\sigma_{\rm f}$ can describe either a classical atomic polarization or the wave-function of a single atomic excitation created by a single photon. \\
%%%%%%%%%%%%%%%%%%Principle of CRIB%%%%%%%%%%%%%%%%%%%%%%%%
The principle of a memory based on reversible absorption can be deduced from the symmetry analysis of the equations of motion. The forward components of the light operator and of the atomic coherence evolve following the set of equations (\ref{atomforward_motion})-(\ref{pulseforward_motion}).
If the following transformations are performed
\begin{subequations}
\begin{equation}
\label{detuning_eq}
\Delta \rightarrow -\Delta,
\end{equation}
\begin{equation}
 E_{\rm b} \rightarrow -E_{b},
\end{equation}
\end{subequations}
the equations for the backward components (\ref{atombackward_motion})-(\ref{pulsebackward_motion}) define a time reversal evolution of the system compared to the equations for the forward components. This analysis reveals the essence of a quantum memory based on CRIB \cite{Moiseev01, Kraus06}: a light pulse propagating in the forward direction is completely absorbed by the atomic ensemble. What is required to retrieve the pulse as a time reversed copy of itself is to reverse the detuning between the spectral components of the light pulse and the atomic transition frequency. At the same time, one has to apply a phase matching operation such that the wave function of the single atomic excitation propagates in the backward direction.\\
We now describe how one can construct an optical memory according to the principle discussed above. The following scenario closely follows the ideas presented in Refs. \cite{Moiseev01, Kraus06}.
\begin{figure}[hr!]
\rotatebox{89.8}
{\includegraphics[scale=0.28]{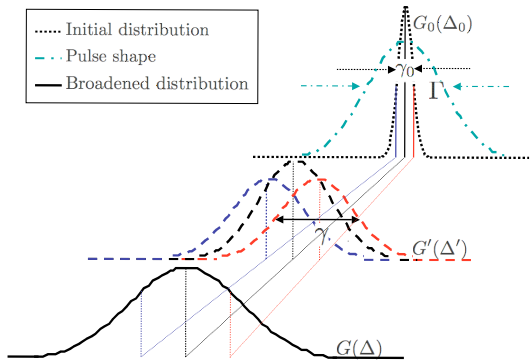}}
\caption{(Color online) Schematic representation of the spectral atomic distribution. The initial distribution $G_0(\Delta_0)$ with characteristic bandwidth $\gamma_0$ is represented as black dotted line. Three spectral components of the initial distribution are considered (blue, black and red vertical lines) and each of them is broadened according to a distribution $G'(\Delta')$ (blue, black and red dashed lines) with bandwidth $\gamma.$ The final broadened distribution, called $G(\Delta)$ is the convolution of the initial distribution $G_0(\Delta_0)$ and of the broadened distribution $G'(\Delta')$ associated to a single initial absorption line $(\Delta_0).$ The pulse shape with bandwidth $\Gamma$ is represented as green dashed-dotted line.}
\label{fig_notation}
\end{figure}

(i) One first chooses the medium. The relevant atomic transition should have a frequency in the optical domain. It should also have low decoherence rate compared to the total duration of the pulse in order to keep the phase and amplitude informations during the optical absorption. In Ref. \cite{Kraus06}, several experimental realizations are discussed using atomic gases at room temperature or ion-doped solid-state materials cooled to cryogenic temperature. Motivated by the work done in our group \cite{Hastings06, Staudt06}, we here consider a medium based on ion-doped solid materials.  \\
(ii) A narrow atomic absorption line is prepared. In ion-doped materials, the absorption transition is broadened inhomogeneously due to the environment in the host which is different for each ion. A narrow absorption line can be prepared using optical pumping techniques. The width of the prepared line is limited ultimately by the homogeneous line-width defined from the decoherence rate of the atomic ensemble. In practice however, the bandwidth of the absorption line might be limited by the bandwidth of the laser used for the optical pumping. We will denote the initial detuning by $\Delta_0$. The selected inhomogeneous atomic distribution, called $G_0(\Delta_0)$ with bandwidth $\gamma_0$ is represented in Fig. {\ref{fig_notation}} as dotted line.\\
(iii) The initial distribution $G_0(\Delta_0)$ associated to the detunings $\Delta_0$ is then broadened to a larger bandwidth using reversible and controlled inhomogeneous broadening. 
Every detuning $\Delta_0$ is broadened to $\Delta_0+\Delta',$ where $\Delta'$ is distributed according to $G'(\Delta')$ (dashed line in Fig. \ref{fig_notation}). In ion-doped materials, a controlled inhomogeneous broadening of the initial absorption line can be obtained by applying an electric field gradient that shifts the transition frequency of the ions by the Stark effect such that the detuning $\Delta'$ is given by the induced Stark shift \cite{Alexander06, Nilsson05, Hastings06, Hetet06}. The final distribution $G(\Delta)$ (full line in Fig. \ref{fig_notation}), taking into account the broadening of all the detunings $\Delta_0,$ is given by the convolution of the initial distribution $G_0(\Delta_0)$ and of the distribution $G'(\Delta').$ By reversing the electric field, the Stark shift is reversed and consequently, each detuning is reversed such that Eqn. (\ref{detuning_eq}) is replaced by
\begin{equation}
\Delta_0+\Delta' \rightarrow \Delta_0-\Delta'.
\end{equation}
Here we consider a controlled inhomogeneous broadening which is independent of the position $z.$ This corresponds to a rare-earth doped fiber or to a crystal in which the electric field gradient is applied in a direction transverse to the propagating pulse.\\
(iv) The equations of motion associated to the forward components are connected to the ones associated to the backward components by applying a position-dependent phase $e^{2i\omega_0z/c}$ to the atomic ensemble when the light excitation is reduced to zero. This transforms the forward component $\sigma_{\rm{f}}$ into the backward component $\sigma_{\rm{b}}$ according to the Eqn. (\ref{sigma}). The phase shift $e^{2i\omega_0z/c}$ can be realized by using counter-propagating $\pi$-pulses to transfer the atomic coherence back and forth to an auxiliary ground state \cite{Nilsson05, Kraus06}. This procedure can also offer longer storage duration by choosing a ground state with long coherence time \cite{Longdell05}. If the phase shift is not applied, the evolution of the system is given by the forward set of equations. The pulse reemitted by the medium propagates in the forward direction and is partially reabsorbed, limiting the memory efficiency. On the other hand, this simplified protocol does not require the use of additional control laser fields. This provides an advantage for few photons quantum memory by avoiding the light noise generated by auxiliary laser fields. \\

In the next section, we study the properties of the optical memory for a finite optical depth when the atomic ensemble reemits the light pulse in forward and backward directions. We also look at the impact of the initial distribution bandwidth and we establish the optimal broadening of the initial distribution with respect to the memory efficiency for a given pulse.
%%%%%%%%%%%%%%%%%%%%%%%%%%%%%%%%%%%%%%%%%%%%%%%%%%
\section{General solution}\label{resolution}
We solve the equations of motion by generalizing the technique presented in ref. \cite{Crisp70}. To clearly distinguish the contribution of the initial distribution of atoms $G_0(\Delta_0)$ and of the broadened distribution $G'(\Delta')$ associated to a single absorption line, we introduce the atomic operator
\begin{equation}
\sigma_{\rm ij}(z,t;\Delta',\Delta_0):= \frac{1}{N(\Delta'+\Delta_0,z)} \sum_{n=1}^{N(\Delta'+\Delta_0,z)} \ket{i}_{nn} \bra{j}|
\end{equation}
where $N(\Delta'+\Delta_0,z)$ is the number of atoms with initial detuning around $\Delta_0$ broadened to $\Delta'+\Delta_0$ and with position around $z.$ When the inhomogeneous broadening is reversed, the detuning is changed from $\Delta'+\Delta_0$ to $-\Delta'+\Delta_0$ such that the atomic properties are described by the operator $\sigma_{\rm ij}(z,t;-\Delta',\Delta_0).$ As in the previous section, the atomic operator is decomposed into two counter-propagating contributions. We are interested in the evolution of an incoming light pulse propagating in the forward direction through the atomic ensemble
\begin{subequations}
\label{set_forward}
\begin{eqnarray}
\label{eqn_field_forward2}
&&\Big(\frac{\partial}{\partial t} + c\frac{\partial}{\partial z}\Big)E_{\rm{f}}^{\rm{in}}(z,t)= \\\nonumber
&&i\beta \int_{-\infty}^{+\infty} d\Delta_0  d\Delta' G(\Delta_0)G'(\Delta') \sigma_{\rm f}(z,t;\Delta',\Delta_0),
\end{eqnarray}
\begin{eqnarray}
\label{eqn_atom_forward2}
&& \frac{\partial}{\partial t} {\sigma}_{\rm f}(z,t;\Delta',\Delta_0)=\\ \nonumber
& &\quad \quad\quad-i(\Delta_0+\Delta') \sigma_{\rm f}(z,t;\Delta',\Delta_0)+i\wp E_{\rm f}^{\rm{in}}(z,t),
\end{eqnarray}
\end{subequations}
Initially, there is no atomic excitation such that  $\sigma_{\rm f}(z,t\rightarrow -\infty;\Delta',\Delta_0)=0.$ Taking into account this initial condition, Eqn. (\ref{eqn_atom_forward2}) can be solved as
\begin{equation}
\label{eqn_sigma_ge}
\sigma_{\rm f}(z,t;\Delta',\Delta_0)=i \wp \int_{-\infty}^{t}ds \-\ e^{-i(\Delta'+\Delta_0)(t-s)} E_{\rm f}^{\rm in}(z,s).
\end{equation}
We are interested in the regime for which the storage duration $T$ is longer than the pulse duration $\tau.$ We thus consider that at the position $z=0,$ the incoming light pulse $E^{\rm{in}}(z=0)$ is centered around $-T/2.$ The inhomogeneous broadening is reversed at time $t=0$ after a time sufficiently long with respect to the pulse duration $\tau.$ The Fourier transform of the incoming light pulse is defined by
\begin{equation}
\tilde{E}_{\rm f}^{\rm{in}}(z,\omega):= \int_{-\infty}^{0} dt \-\ e^{i\omega t} E_{\rm f}^{\rm{in}}(z,t).
\end{equation}
The upper bound is equal to $0$ since the incoming light pulse is not defined anymore at positive times. Furthermore, we consider the regime in which the broadened distribution bandwidth $\gamma$ is larger than the inverse of the storage duration $1/T.$ These considerations allow us to introduce the Fourier transform of the incoming light pulse in Eqn. (\ref{eqn_field_forward2}) in which $\sigma_{\rm f}$ is replaced by its expression (\ref{eqn_sigma_ge})
\begin{equation}
\label{propag_tf}
\left(\frac{\partial}{\partial z}-\frac{i\omega}{c}+\eta H(\omega)\right)\tilde{E}_{\rm f}^{\rm{in}}(z,\omega)=0.
\end{equation}
$H(\omega)$ is defined by
\begin{eqnarray}
\nonumber
H(\omega)&:= &\int_{0}^{+\infty} dx \Big(e^{i\omega x} \times\\
&& \nonumber \int_{-\infty}^{+\infty} d\Delta_0 d\Delta' G_0(\Delta_0) G'(\Delta') e^{-i(\Delta'+\Delta_0) x}\Big)
\end{eqnarray}
and $\eta:=g_0^2N\wp^2/c.$
The solution of Eqn. (\ref{propag_tf}) is given by
\begin{equation}
\label{sol_field_firstset}
\tilde{E}_{\rm f}^{\rm{in}}(z,\omega)=\tilde{E}_{\rm f}^{\rm{in}}(0,\omega) e^{i\omega z/c}e^{-\eta H(\omega) z}
\end{equation}
at any position inside the medium $0\leqslant z \leqslant L.$
The solutions (\ref{eqn_sigma_ge}) and (\ref{sol_field_firstset}) describes the absorption of the incoming light pulse by the atomic ensemble.
The absorption coefficient $\alpha(\omega)$ defined from the relation
$
|\tilde{E}_{\rm f}^{\rm{in}}(z,\omega)|^2=e^{-\alpha(\omega)z}|\tilde{E}_{\rm f}^{\rm{in}}(0,\omega)|^2
$
is given by
\begin{equation}
\label{absorption_coefficient}
\alpha(\omega):=\frac{2g_0^2N \wp^2 \rm{Re}(H(\omega))}{c}.
\end{equation}
It is thus proportional to the coupling between the light pulse and the atomic ensemble and depends on the atomic distribution via the function $H(\omega).$ \\
In what follows, we give the expressions of the outgoing light pulse reemitted by the atomic ensemble by distinguishing the pulses reemitted in the controlled backward  and forward directions.

%%%%%%%%%%%%%%%%%%%%%%%%%%%%%%%%%%%%%%%%%%%%%%%%%%
\subsection{Complete protocol: Emission in backward direction}
The light pulse is partially absorbed such that a part of the light excitation is transfered into the atomic ensemble. At time $t=0$, the inhomogeneous broadening is reversed meaning that the detuning $\Delta_0+\Delta'$ changes to $\Delta_0-\Delta'.$ Furthermore, the phase shift $e^{2i\omega_0z/c}$ is applied to the atomic ensemble
such that the system evolves backward in space. We assume that at $t=0$ the non-absorbed part of the light has left the medium so that the field is zero, i.e. the excitation is purely atomic. The system is described by the following set of equations
\begin{subequations}
\begin{eqnarray}
\label{eqn_pulse_backward2}
&&\left(\frac{\partial}{\partial t} - c\frac{\partial}{\partial z}\right) E_{\rm{b}}^{\rm{out}}(z,t)=\\
& & i\beta\int_{-\infty}^{+\infty} d\Delta_0 d\Delta' G_0(\Delta_0)G(-\Delta') \sigma_{\rm b}(z,t;-\Delta',\Delta_0),\nonumber
\end{eqnarray}
\begin{eqnarray}
\label{eqn_atom_backward2}
&&\frac{\partial}{\partial t}{\sigma}_{\rm b}(z,t;-\Delta',\Delta_0)=\\
& & \quad \quad \quad  -i(\Delta_0-\Delta') \sigma_{\rm b}(z,t;-\Delta',\Delta_0)+i\wp E_{\rm b}^{\rm{out}}(z,t)
\nonumber
\end{eqnarray}
\end{subequations}
with the initial condition
\begin{eqnarray}
&&\nonumber E_{\rm{b}}^{\rm{out}}(z,t=0)=0, \\
&& \nonumber \sigma_{\rm b}(z,t=0;-\Delta',\Delta_0)= i\wp \int_{-\infty}^{0} ds \-\ e^{i(\Delta'+\Delta_0) s} E_{\rm f}^{\rm{in}}(z,s).
\end{eqnarray}
The solution of Eqn. (\ref{eqn_atom_backward2}) is given by
\begin{eqnarray}
&\nonumber \sigma_{\rm b}(z,t;-\Delta',\Delta_0)=\sigma_{\rm b}(z,0;-\Delta',\Delta_0)+\int_{0}^{t}ds \big[e^{i(\Delta'-\Delta_0)(t-s)}& \\
&\times\left(i\wp \-\ E_{\rm b}^{\rm{out}}(z,s) + i(\Delta'-\Delta_0) \-\ \sigma_{\rm b}(z,0;-\Delta',\Delta_0)\right)\big].&
\nonumber
\end{eqnarray}
Plugging the previous expression into Eqn. (\ref{eqn_pulse_backward2}), we get the following propagation equation
\begin{eqnarray}
\label{propag_backward}
& &\left(\frac{\partial}{\partial z}+\frac{i\omega}{c}-\eta F(\omega)\right)\tilde{E}_{\rm b}^{\rm{out}}(z,\omega)= \\  \nonumber
&&\eta \int_{-\infty}^{+\infty} d\Delta_0 G_0(\Delta_0) J(\omega;\Delta_0) \tilde{E}_{\rm f}^{\rm in}(z,-\omega+2\Delta_0)
\end{eqnarray}
for the Fourier transform of the outgoing pulse
\begin{equation}
\tilde{E}_{\rm b}^{\rm out}(z,\omega):=\int_{0}^{+\infty} dt \-\ e^{i\omega t} E_{\rm b}^{\rm out}(z,t).
\end{equation}
The lower bound is equal to $0$ since the outgoing light pulse, centered around $T/2,$ is not defined at negative times.
We have introduced the functions
\begin{equation}
\nonumber
J(\omega;\Delta_0):= \int_{-\infty}^{+\infty} dx \int_{-\infty}^{+\infty} d\Delta' G'(-\Delta')e^{i(\Delta'-\Delta_0)x} e^{i\omega x}
\end{equation}
and
\begin{eqnarray}
&\nonumber F(\omega):=& \int_{0}^{+\infty} dx \Big(e^{i\omega x} \times \\
&&\nonumber \int_{-\infty}^{+\infty} d\Delta_0 d\Delta' G_0(\Delta_0)G'(-\Delta')e^{i(\Delta'-\Delta_0)x}\Big).
\end{eqnarray}
The solution of Eqn. (\ref{propag_backward}) establishes the connection between the output and the input light pulses when they propagate in opposed directions
\begin{eqnarray}
\label{exp_BackwardField}
&&\nonumber \tilde{E}_{\rm b}^{\rm out}(z,\omega)=-\eta \int_{-\infty}^{+\infty} d\Delta_0 G_0(\Delta_0) e^{-i\omega z/c} \times \\
&&\nonumber \Big[e^{2i\Delta_0L/c}e^{-\eta L H(-\omega+2\Delta_0)}e^{\eta(z-L)F(\omega)}-\\
&&\nonumber e^{2i\Delta_0z/c}e^{-\eta z H(-\omega+2\Delta_0)} \Big] \times\\
&&\nonumber  \frac{J(\omega;\Delta_0)}{2i\Delta_0/c-\eta H(-\omega+2\Delta_0)-\eta F(\omega)} \tilde{E}_{\rm f}^{\rm{in}}(0,-\omega+2\Delta_0).\\
\end{eqnarray}
In the next subsection, we study the simplified protocol in which the position-dependent phase is not applied to the atomic ensemble. We find the expression of the output pulse when it propagates in the forward direction.
%%%%%%%%%%%%%%%%%%%%%%%%%%%%%%%%%%%%%%%%%%%%%%%%%%
\subsection{Protocol without phase shift: Emission in forward direction}\label{forward_crib_sub}
If at time $t=0,$ the additional phase $e^{2iw_0z/c}$ is not applied to the atomic ensemble, the system evolves forward in space and its evolution is given by the set of equations (\ref{set_forward}) in which $\Delta'$ has to be changed to $-\Delta'.$ As above, the light pulse is partially absorbed and the absorbed light excitation is transfered to the atoms. The initial condition is thus
\begin{eqnarray}
&&\nonumber E_{\rm{f}}^{\rm{out}}(z,t=0)=0, \\
&& \nonumber \sigma_{\rm f}(z,t=0;-\Delta',\Delta_0)= i\wp \int_{-\infty}^{0} ds \-\ e^{i(\Delta'+\Delta_0) s} E_{\rm f}^{\rm{in}}(z,s).
\end{eqnarray}
The equations of motion are resolved as previously. The expression of the outgoing light pulse propagating in the forward direction is given by
\begin{eqnarray}
\label{exp_ForwardField}
&&\nonumber \tilde{E}_{\rm f}^{\rm out}(z,\omega)=-\eta z \int_{-\infty}^{+\infty} d\Delta_0 G_0(\Delta_0)J(\omega;\Delta_0) \times \\ \nonumber
&& {\rm{sinhc}}\left(\eta z\frac{F(\omega)-H(-\omega+2\Delta_0)}{2}-iz\frac{\omega-\Delta_0}{c}\right) \times \\ \nonumber
&&\exp \left({iz\frac{\Delta_0}{c}-\eta z\frac{F(\omega)+H(-\omega+2\Delta_0)}{2}}\right) \times \\
&&\tilde{E}_{\rm f}^{\rm in}(0,-\omega+2\Delta_0).
\end{eqnarray}
The sinhc denotes the hyperbolic sinus cardinal function $({\rm{sinhc}}(x)=\sinh(x)/x).$\\
 The equations (\ref{exp_BackwardField}) and (\ref{exp_ForwardField}) establish the expression of the light pulse reemitted by the atomic ensemble in both backward and forward directions as a function of the input light pulse characteristics and of the properties of the atomic medium. They are valid for a general initial distribution broadened to an arbitrary distribution. In the next sections, they are analyzed in order to  deduce some characteristics of the memory for specific atomic distributions.
%%%%%%%%%%%%%%%%%%%%%%%%%%%%%%%%%%%%%%%%%%%%%%%%%%%
\section{Finite optical depth}\label{finite_optical_depth}
%%%%%%%%%%%%%%%%%%%%%%%%%%%%%%%%%%%%%%%%%%%%%%%%%%%
In the framework of a practical realization, it is necessary to know the properties of the memory for a finite optical depth. This topic has also been addressed for the CRIB protocol in Refs. \cite{Moiseev04,Gorshkov3} and for the slow-light based memories in Refs. \cite{Gorshkov1, Gorshkov2}. \\

\begin{figure}[hr!]
\rotatebox{89.8}{
\includegraphics[scale=0.25]{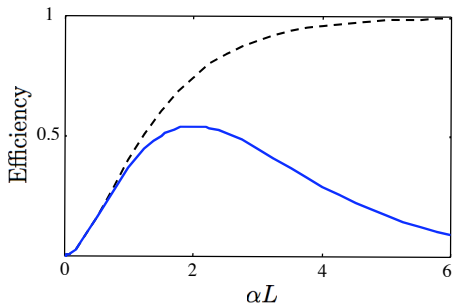}}
\caption{(Color online) Efficiency of the light pulse reemission in forward (full blue line) and backward (dashed black line) direction as a function of the optical depth for a constant broadening. The efficiencies vary as $(\alpha L)^2 e^{-\alpha L}$ for the forward protocol and as $(1-e^{-\alpha L})^2$ for the backward protocol.}\label{fig1}
\end{figure}

The analysis is done using a simple model in which the initial distribution is reduced to a single narrow absorption line $G_0(\Delta_0)=\delta(\Delta_0).$ The initial absorption line is broadened to be constant on the interval $[-\gamma/2,\gamma/2]$ and null elsewhere such that it satisfies the condition
$
\int_{-\infty}^{+\infty} d\Delta' G'(\Delta')=\int_{-\infty}^{+\infty} d\Delta G(\Delta)=1.
$
In the regime where the spectral pulse bandwidth is smaller than the inhomogeneous broadening $\Gamma \ll \gamma,$ we have
$
H(\omega)\approx F(\omega)\approx J(\omega;0)/2 \approx \pi/\gamma
$
and the outgoing pulse propagating in the backward direction (\ref{exp_BackwardField}) takes the following form
\begin{equation}
\label{exp_pulse_delta01}
E_{\rm b}^{\rm{out}}(0,t)= -(1- e^{-\alpha L}) E_{\rm f}^{\rm{in}}(0,-t).
\end{equation}
at the position $z=0.$ The absorption coefficient is given by
$
\alpha=2\pi g_0^2N \wp^2/\gamma c.
$
The memory efficiency, defined by
\begin{equation}
\label{Efficiency}
\rm{Eff}:=\frac{\int d\omega |\tilde{E}^{\rm out}(\omega)|^2}{\int d\omega |\tilde{E}^{\rm in}(\omega)|^2},
\end{equation}
corresponds to the probability to retrieve an absorbed photon. The memory efficiency depends on the optical depth $\alpha L$ and as can be seen in Fig. \ref{fig1}, it reaches 100\% for large enough optical depth. This limit of a large optical depth corresponds to the result presented in Ref. \cite{Kraus06}.

Under similar conditions, when no phase shift is applied, the expression of the output light pulse propagating in the forward direction (\ref{exp_ForwardField}) is given at the position $z=L$ by
\begin{equation}
\tilde{E}_{\rm f}^{\rm{out}}(L,\omega)=-\alpha L e^{-\alpha L/2}
\frac{\sin\left(\omega L/c\right)}{\omega L/c} \tilde{E}_{\rm f}^{\rm{in}}(0,-\omega).
\end{equation}

The sinus cardinal function induces a distortion of the light for large pulse bandwidth $\Gamma$ with respect to $c/L.$ For a pulse bandwidth smaller than $c/L,$ the expression for the forward pulse is reduced to
\begin{equation}
\label{exp_pulse_delta0}
E_{\rm f}^{\rm{out}}(L,t)=-\alpha L e^{-\alpha L/2} E_{\rm f}^{\rm{in}}(0,-t).
\end{equation}
In this regime, the interaction with the atomic ensemble does not induce distortion neither in the forward direction nor in the backward direction. For small optical depth, the memory efficiency varies quadratically as a function of the optical depth for both forward and backward protocols. However, in the forward direction, the pulse is reabsorbed by the atoms when the optical depth increases and the memory efficiency reaches a maximum of 54\% at the optical depth $\alpha L=2.$ Here we have supposed that the optical depth $\alpha L$ is constant for the whole spectral width of the pulse. If $\alpha L$ is lower for certain spectral components of the pulses, the memory efficiency is lower.
%%%%%%%%%%%%%%%%%%%%%%%%%%%%%%%%%%%%%%%%%%%%%%%%%%%
\section{Atomic distribution}\label{discussion}
In the previous section, we have studied the dependence of the memory efficiency on the optical depth. For an initial distribution with characteristic bandwidth $G_0(\Delta_0) \approx \gamma_0$ smaller than the spectral pulse bandwidth $\Gamma$ and broadened to a distribution bandwidth $G(\Delta)\approx \gamma,$ the optical depth can be written as
\begin{equation}
\label{optical_depth}
\alpha L \approx \frac{g_0^2 \wp^2 \rho_0(0) L}{c} \frac{\gamma_0}{\gamma}=\alpha_0 L \frac{\gamma_0}{\gamma}
\end{equation}
where $\rho_0(\Delta_0)=N G_0(\Delta_0)$ is the initial density of atoms. $\alpha_0L$ corresponds to the initial optical depth in the absence of broadening. For a given medium , the optical depth $\alpha L$ varies with the ratio of the initial distribution bandwidth on the broadened distribution bandwidth. In a first  subsection, we show that the initial distribution bandwidth also limits the storage duration. We thus prove the existence of a trade-off between the memory efficiency and the storage duration. In a second subsection, we look at the optimal broadening for a given pulse and for a fixed initial distribution.
%%%%%%%%%%%%%%%%%%%%%%%%%%%%%%%%%%%%%%%%%%%%%%%%%%%
\subsection{Memory efficiency versus storage duration}\label{memo_eff}
To analyze the role of the initial distribution bandwidth, we consider the simple case in which the broadened distribution is constant in a given interval and null elsewhere
\begin{equation}
G'(\Delta')=\frac{1}{\gamma} \theta(\Delta'+\frac{\gamma}{2})\theta(\frac{\gamma}{2}-\Delta')
\end{equation}
($\theta$ being the Heaviside function) such that in the regime $\Gamma \ll \gamma$ and $\gamma_0 \ll \gamma,$ we have $J(\omega,\Delta_0)/2 \approx H(\omega) \approx F(\omega)\approx \pi/\gamma.$ We discuss the results obtained from the backward process but the conclusions are applicable for the forward process as well. Under the condition $\gamma_0 \ll c/L,$ the output light pulse emitted in the backward direction (\ref{exp_BackwardField}) is given by the following expression
\begin{eqnarray}
\label{output_pulse_delta0}
&&E_{\rm b}^{\rm{out}}(0,t)=-(1-e^{-\alpha L}) E_{\rm f}^{\rm{in}}(0,-t) \times \nonumber \\
& & \int_{-\infty}^{+\infty} d\Delta_0 G_0(\Delta_0) e^{-2i\Delta_0 t}
\end{eqnarray}
at the position $z=0.$ Since the output pulse is centered around $T/2,$ we are interested in times $t$ around $T/2.$ Comparing Eqns. (\ref{exp_pulse_delta01}) and (\ref{output_pulse_delta0}), it clearly appears that the output pulse is multiplied by the Fourier transform of the initial distribution. For an initial distribution bandwidth $\gamma_0,$ the storage duration $T$ is thus limited by $1/\gamma_0.$ This has already been observed experimentally as reported in ref. \cite{Alexander06}. The bandwidth of the initial distribution thus constitutes a limitation for the storage duration $T\gamma_0 \ll 1.$ Since the initial distribution width also affects the optical depth (see Eqn. (\ref{optical_depth})) and thus the memory efficiency, there exits a trade-off between a long storage duration and a large memory efficiency. Note that this is true for a two-level system. If other long-living levels are available, the storage time can be increased significantly by transferring the excitation to a second long-living state, e.g. a ground state hyperfine level \cite{Longdell05}. 
%%%%%%%%%%%%%%%%%%%%%%%%%%%%%%%%%%%%%%%%%%%%%%%%%%%
\subsection{Optimal broadening}
We now consider that the initial distribution has been chosen. We define the effective width of the initial distribution
\begin{equation}
\nu:=\frac{g_0^2N \wp^2 L}{c} \approx \frac{g_0^2 \wp^2 \rho_0(0) \gamma_0 L}{c} =\alpha_0 L \gamma_0
\end{equation}
which depends on the medium properties together with the initial distribution and thus has a fixed value.
The optical depth after broadening is given by 
\begin{equation}
\label{alphaL}
\alpha L \approx \nu/\gamma.
\end{equation}
For a given pulse shape, we look for the optimal width of the broadened distribution with respect to the memory efficiency.\\

%%%%%%%%%%%%%%%%%%%%%%%%%%%%%%%%%%%%%%%%%%%%%%%%%%%
We first fix the shape of the atomic distributions to be the same as the Lorentzian shape of the light pulse
\begin{eqnarray}
\label{Lorentzian_pulse}
&& \tilde{E}_{\rm f}^{\rm{in}}(0,\omega)=\frac{\Gamma}{2\pi}\frac{1}{\Gamma^2/4+\omega^2}\\
\label{Lorentzian_initial_distribution}
&& G_0(\Delta_0)=\frac{\gamma_0}{2\pi}\frac{1}{\gamma_0^2/4+\Delta_0^2}\\
\label{Lorentzian_broadened_distribution}
&& G'(\Delta')=\frac{\gamma}{2\pi}\frac{1}{\gamma^2/4+\Delta'^2}.
\end{eqnarray}
For a given value of the effective width of the initial distribution $\nu,$ we calculate the memory efficiency by varying the broadened distribution width with respect to the pulse bandwidth. In the limit $\gamma_0 \rightarrow 0,$ $\rho_0(0) \rightarrow +\infty$ such that $\nu$ has a finite value, the functions $J(\omega;0),$ $F(\omega)$ and $H(\omega)$ take the simple forms
\begin{equation}
J(\omega;0)=\frac{\gamma}{\gamma^2/4+\omega^2}, \quad H(\omega)=F(\omega)=\frac{\gamma/2+i\omega}{\gamma^2/4+\omega^2}.
\end{equation}
The output pulse (\ref{exp_BackwardField})-(\ref{exp_ForwardField}) is given by
\begin{equation}
\tilde{E}_{\rm b}^{\rm out}(0,\omega)=-\frac{\Gamma}{2\pi(\Gamma^2/4+\omega^2)}\left(1-\exp{\left(-\frac{\nu\gamma}{\gamma^2/4+\omega^2}\right)} \right),
\end{equation}
\begin{eqnarray}
& &\tilde{E}_{\rm f}^{\rm out}(L,\omega)= -\frac{\Gamma}{2\pi(\Gamma^2/4+\omega^2)}\frac{\nu \gamma}{\gamma^2/4+\omega^2}\times  \nonumber \\
& & \exp{\left(-\frac{\nu \gamma}{2(\gamma^2/4+\omega^2)}\right)} \rm{sinc}\left(\frac{\nu \omega}{\omega^2+\gamma^2}\right),
\end{eqnarray}
when propagating in the backward and forward (with the hypothesis $\Gamma$ smaller than $c/L$) directions respectively. \\
\begin{figure}[ht!]
\rotatebox{89.8}{
\includegraphics[scale=0.4]{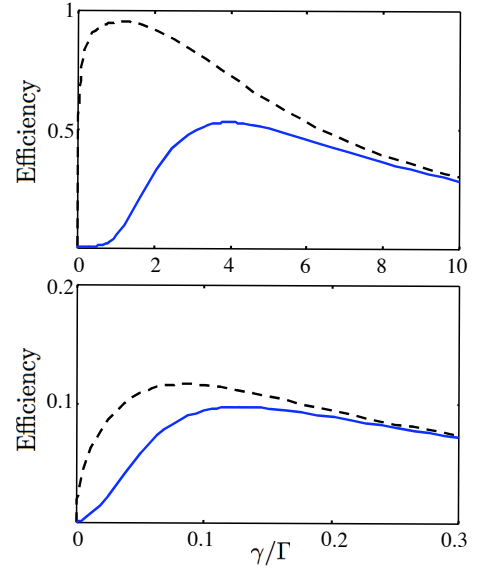}}
\caption{(Color online) Efficiency of the light pulse reemission for forward (full blue line) and backward (dashed black line) protocol as a function of the broadened distribution bandwidth $\gamma$ (in units of the pulse bandwidth $\Gamma$) for effective widths of the initial distribution $\nu=2\Gamma$ (top) and $\nu=0.05\Gamma$ (bottom).}
\label{fig2}
\end{figure}
In Fig. \ref{fig2} (top), we take $\nu=2 \Gamma$ and we plot the efficiency as a function of the broadened distribution bandwidth $\gamma.$ For large $\gamma,$ the optical depth (\ref{optical_depth}) which is inversely proportional to $\gamma$, is small. As shown in Fig. \ref{fig2} (top), the memory efficiency for the backward protocol is close to the efficiency of the forward protocol. For smaller values of $\gamma,$ the efficiency increases and reaches a maximum for the forward process caused by the reabsorption while the efficiency of the backward protocol continues to grow until a value close to 100\%. When $\gamma$ tends to zero, the bandwidth of the broadened distribution is too small to absorb the light pulse and the memory efficiency tends to zero for both backward and forward processes. In both cases, however, the optimum efficiency occurs when the broadened distribution bandwidth and the light bandwidth are of the same order.\\
Taking a smaller value for  $\nu=0.05\Gamma,$ we plot in Fig. \ref{fig2} (bottom) the emission efficiency versus the distribution bandwidth. The efficiency curves exhibit similar shapes as in the large $\nu$ case except that their maxima reach smaller values. These maxima are obtained  for a narrower broadened distribution bandwidth than the pulse bandwidth.\\
One can understand these results in the following way: To insure a high efficiency of the CRIB protocol, the optical depth after broadening has to satisfy $\alpha L \gtrsim 1$. From Eqn. (\ref{alphaL}), we deduce that the effective width of the initial distribution has to be at least of the order of the broadened distribution bandwidth: $\nu \gtrsim \gamma.$ Thus, for a pulse bandwidth $\Gamma \approx \nu,$ one can broaden the initial distribution until $\gamma \approx \Gamma$ and thus achieve high efficiency. If $\Gamma \gg \nu,$ it is more advantageous to choose a smaller broadening than the pulse width $\gamma \approx \nu  \ll \Gamma.$
 In this case, however, one must expect strong distortion of the output pulse due to filtering effects.

%%%%%%%%%%%%%%%%%%%%%%%%%%%%%%%%%%%%%%%%%%%%%%%%%%%
\subsection{Broadening shape}
\begin{figure}[ht!]
\rotatebox{89.8}{
\includegraphics[scale=0.25]{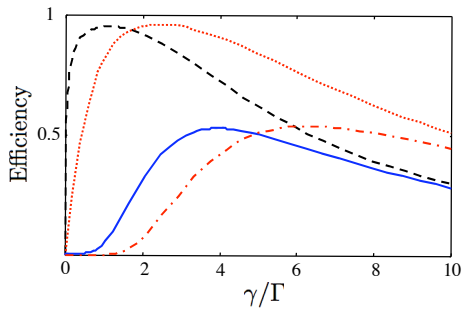}}
\caption{(Color online) Comparison of memory efficiencies for a spectral distribution of atoms and light pulse with identical shapes (forward: full blue line, backward: dashed black line) and with different shapes (forward: dashed-dotted red line, backward: dotted red line) as a function of the inhomogeneous bandwidth (in units of the pulse bandwidth). The effective width of the initial distribution $\nu$ is taken equal to $2\Gamma.$}
\label{fig4}
\end{figure}

Finally, we investigate how the shape of the broadened distribution influences the memory efficiency. We keep the Lorentzian shape of the light pulse but we take a constant distribution in a given interval
\begin{eqnarray}
&& \tilde{E}_{\rm f}^{\rm{in}}(0,\omega)=\frac{\Gamma}{2\pi}\frac{1}{\Gamma^2/4+\omega^2},\\
&& G_0(\Delta_0)=\frac{1}{\gamma_0}\theta(\Delta_0+\gamma_0/2)\theta_0(\gamma_0/2-\Delta_0),\\
&& G'(\Delta')=\frac{1}{\gamma}\theta(\Delta'+\gamma/2)\theta(\gamma/2-\Delta').
\end{eqnarray}
For a fixed value of the effective width of the initial distribution $\nu,$ we calculate the memory efficiency and we compare it to the efficiency obtained for the Lorentzian shapes of Eqns. (\ref{Lorentzian_pulse}-\ref{Lorentzian_broadened_distribution}).\\
In the limit $\gamma_0 \rightarrow 0,$ the functions $J,$ $F$ and $H$ take the following forms
\begin{eqnarray}
&& J(\omega;0)=\frac{2\pi}{\gamma} \theta(\omega+\gamma/2) \theta(\gamma/2-\omega), \\
&& F(\omega)=H(-\omega)^{\ast}=\frac{J(\omega)}{2}+\frac{i}{\gamma} \log\frac{\gamma+2\omega}{\gamma-2\omega},
\end{eqnarray}
such that the light pulse generated in the backward direction (\ref{exp_BackwardField}) is given by
\begin{equation}
\tilde{E}_{\rm b}^{\rm out}(0,\omega)=-\frac{\Gamma}{2\pi}\left(1-e^{-\frac{2\pi\nu}{\gamma}} \right) \frac{1}{\Gamma^2/4+\omega^2}
\end{equation}
for $\omega \in [-\gamma/2,\gamma/2]$ and $0$ elsewhere. For $\Gamma \ll c/L,$ the pulse generated in the forward direction (\ref{exp_ForwardField}) is of the form
\begin{equation}
\tilde{E}_{\rm f}^{\rm out}(L,\omega) \approx -\frac{\nu}{\gamma} e^{-\frac{\pi\nu}{\gamma}} \frac{\Gamma}{\Gamma^2/4+\omega^2}
\end{equation}
for $\omega \in [-\gamma/2,\gamma/2].$ From these expressions, we plot (see Fig. \ref{fig4}) the memory efficiency. We compare it to the efficiency we found for identical broadened distribution and light pulse shapes. Globally, the relative shapes of the broadened distribution and of the light pulse does not modify  the memory efficiency. Locally, for small distribution bandwidths, the memory efficiency is a bit larger for identical shapes. However, when the distribution bandwidth increases, the efficiency is higher when these shapes are different.
%%%%%%%%%%%%%%%%%%%%%%%%%%%%%%%%%%%%%%%%%%%%%%%%%%%
\section{Conclusion}
\label{conclusion}
We have obtained the explicit solution of the equations of
motion for the system atoms plus light in the weak excitation regime, making it possible to
gain insight into the dependence of the memory efficiency on the
optical depth, and on the width and shape of the atomic
spectral distributions. Furthermore, we introduced a simplified CRIB-based memory protocol, which does not require the use of any additional laser field, and showed that its efficiency is limited to 54\%. For both the complete and the simplified CRIB protocols, the interaction with the atoms does not induce distortion of the stored light pulse. We have shown that the storage duration is limited by the initial spectral distribution of the atoms. This distribution also defines the absorption of the medium which determines the memory efficiency. There thus exists a trade-off between the storage duration and the memory efficiency. For a given pulse shape, the optimal broadening of the initial distribution is of the order of the \textit{effective bandwidth} of the initial atomic distribution, defined as the product of the atomic density and the width of the initial atomic distribution. We have also shown that the shape of the broadened distribution does not dramatically change the memory efficiency. \\

%%%%%%%%%%%%%%%%%%%%%%%%%%%%%%%%%%%%%%%%%%%%%%%%
\begin{acknowledgments}
The authors thank S. Hastings-Simon, S.A. Moiseev, V. Scarani  M. Staudt and W. Tittel for enlightening discussions and useful comments. We acknowledge support from the EU integrated project Quantum applications (QAP) and the Swiss NCCR Quantum Photonics.
\end{acknowledgments}

%%%%%%%%%%%%%%%%%%%%%%%%%%%%%%%%%%%%%%%%%%%%%%%%%%%


\begin{thebibliography}{100}
\bibitem{symmetry_density} We assume no particular symmetry for $\rho(\Delta).$ Its form does not play any role for the general time reversal argument. However, it influences the fidelity of the memory in case of weak absorption.
\end{thebibliography}
\end{document}